# An Information-Theoretical Approach to the Information Capacity and Cost-Effectiveness Evaluation of Color Palettes

**Refik Tanju Sirmen**

Istanbul Technical University, Institute of Science, Engineering & Tech.
Turkey

**Burak Berk Üstündağ**

Istanbul Technical University, Faculty of Computer & Informatics
Turkey

## Abstract

Colors are used as effective tools of representing and transferring information. Number of colors in a palette is the direct arbiter of the information conveying capacity. Yet it should be well elaborated, since increasing the entropy by adding colors comes with its cost on decoding. Despite the possible effects upon diverse applications, a methodology for cost-effectiveness evaluation of palettes seems deficient. In this work, this need is being addressed from an information-theoretical perspective, via the articulated metrics and formulae. Besides, the proposed metrics are computed for some developed and known palettes, and observed results are evaluated.

## 1 Introduction

A typical encoder produces symbols representing the message to be transferred, with the generic goal of optimizing the information conveying capacity. Among various sufficiently strong representative instruments, here particularly *colors* are considered. Since the ultimate issue at the receiving end is the decoding authenticity, consistently discriminating colors is the basic problem. Aside from the construction quality of the color palette, theoretical limits of the information conveying capacity is set by the number of colors included. The development of coding schemes with capacities close to its limits is one of the main goals of the information theory.

Comparing palettes of the same number of colors essentially helps choosing the best palette, as a part of the construction. Whereas comparing palettes of different sizes better attends deciding the number of colors to be included. For being able to evaluate the cost-effectiveness of adding colors, having measurable and plausible metrics is imperative. Although it is obvious, this issue seems open. In this dissertation, metrics for this need are explicitly defined, and applied on some developed and known palettes. The *color-difference* measure in the RGB color space is assumed to be the basis of design and discrimination. Also, among diverse areas, barcode applications are taken as the exemplification, because of their wide-spread use, high payloads and archetypal features. However, the offered metrics are generalized as being independent of the considered number of colors, the encoding and decoding system, the transferring and receiving method, the application area, or such. Thus, by following the essential approach, it is possible to derive equivalences for different color spaces and distance criteria.

Here after depicting the problem definition and scope, next the related studies are briefed. In the following section, discussions on the issue around the concept of entropy would provide the necessary work ground. Then the suggested metrics are introduced under the related titles. The derived cost-effectiveness metrics are computed and compared for some developed and existing palettes, and the outcomes are shared under the results. Later these comparisons are commented, and observations are discussed. Conclusion section highlights the major contributions attained.

Improving the information conveying capacity of codes is a generic objective, having impacts upon a multitude of diverse application areas. While the first primary goal of the information theory is the development of fundamental theoretical limits on the achievable performance; the second is the development of coding schemes providing performance reasonably good [4]. Hence, works pertaining these objectives are extensive. However, studies devoted explicitly on the information capacity of color codes are comparatively scarce.

There exist numerous studies and patents on utilizing colors in barcodes, as early as 1972; while many of them are on proposing methods for selecting and calibrating colors. In [6], 10 experimentally selected colors are namely advised. Announced in 2009, the HCCB of Microsoft uses 5 or 8 colors [5]. Quite a few being empirical, the rate of capacity gains, even the justification of the color selection are frequently omitted. Although mentions the attained information capacity, the material of the HCCB supplies almost no discussions on basis of the palette sizes, nor the selection. Some studies focus essentially on the algorithms for decoding the color information. A discussion on the information rate (and its maximum achievable) can be found in [1]. Their experimental sets were selected from a checkerboard of colors, uniformly sampled in RGB space. While one of the goals is pointed out as to encode information with high spatial density, this work rather tries developing a new approach for reliable fast decoding on a smart phone. Also referring to information density, printing of two colors at the same spatial location so that utilizing the separability of colorants is studied in [3]. It is claimed that, this method achieves a significantly higher embedding rate. A colored QR code design is worked in [7]. They claim that their 8-colored version can reach data densities very close to the values reported for HCCB.

Despite the common intuitive conjecture of enlarging the palette size would increase the information capacity, works especially on deciding the optimum number of colors are absent. Similarly, specific suggestions of methods or metrics with plausible generalization for comparing palettes seem scrimpy. The discussions and the suggested metric in this study is expected to serve the field, since it could help development of color palettes of higher information capacities.

## 2 Information capacity of color symbols

At the beginning, harmonizing the terminology with the theory would be useful. As the definitions suggest in [4] "an information source .. is a mathematical model for a physical entity that produces a succession of symbols called "outputs" in a random manner." The space of all of the possible symbols is the alphabet of the source. A source is a probability assignment to events which are sets of symbol sequences from the alphabet. All quantities of a random process naturally associate with the underlying probability space [4]. In the context, the message is encoded through the rules of a *symbology*, via the symbols of an *alphabet*. A *symbol* itself, is single or multi-colored mark of a particular *pattern*. Representative notation of a symbol *color* bases on the color space being worked in. If the RGB space is referenced, then it is denoted as a 3 dimensional vector of primaries $c_i(r,g,b)$. Thus, the colors selected from the space constitute the *palette*. Just like the letters of a formal language, symbol patterns may also take various forms from a finite set of discrete figures, which may be called a pattern *album*. Representation of symbol patterns is not of our concern for the moment; yet the approach is intuitively expected to be analogous. The encoding of the message into symbols -which are from the finite space of the *alphabet*- is supposed to be a random process described by its underlying probability distribution function.

### 2.1 Entropy of symbols

Authenticating symbol patterns and colors are somehow congeneric. In a single-color symbology, once the mark read is deemed as a symbol, then the problem is mapping the detected pattern to the predefined *album space*. Likewise, if the mark is colored, similarly the challenge is to map the perceived color to the predefined *palette space*. Under certain assumptions, including colors and / or patterns naturally enlarges the alphabet space, so then the amount of uncertainty in an individual symbol. Before exploring this any further, depicting these assumptions and resetting the problem domain would be necessary. About the symbology under examination we can reasonably assume the following properties: 1) Each and every symbol color or pattern is equiprobable in their own spaces. 2) Every pattern of the album can be printed in every color of the palette. 3) It is a memoryless, non-Markov source; thus each symbol is independent of all previous ones with the same probability distribution. 4) Encoding is an ergodic process with an underlying uniform random distribution function.

The distribution of the encoded symbols yields being asymptotically uniform, i.e. all symbols are equally likely. We are dealing with an ergodic process, in which every perception or sizable event sample is equally representative of the whole as in regard to some statistical parameter. This assumption enables us "to identify averages along a sequence with averages over the ensemble of possible sequences (the probability of a discrepancy being zero)" as Shannon suggests in [8].

Now, if it is a multi-color single-pattern symbology, and $N_c$ is the number of colors in the palette, regarding to the properties above, the occurrence probability of each color $p(c_i)$ would be $p(c_i) = 1 / N_c$. Then the information revealed by the occurrence of a certain color $c_i$ is $I(c_i) = log (1 / p(c_i)) = log N_c$.

Since all $p(c_i)$ are equal, then the average amount of information is a monotonic increasing function of the number of sample events. Thus, the entropy per symbol *H($S_c$) bits / symbol* would be:

$$H(S_c) = \sum_{i=0}^{N_c} p(c_i)I(c_i) = \sum 1/N_c \, logN_c = N_c logN_c \tag{1}$$

For a single-color multi-pattern symbology, the following could be derived also in the same manner ($N_p$ is the number of patterns): $H(S_p) = N_p logN_p$

Introducing $N_p$ patterns to $N_c$ colors (holding with the 2nd property assumption above) increases the symbology alphabet size to $N_c \times N_p$. Then the entropy:

$$H(S_{cp}) = N_c \times N_p log(N_c \times N_p) \tag{2}$$

These are derived under the mentioned assumptions, yet if the likelihood of any color or pattern (by itself, or in relation) is diverse, the need of a reformulation would become obvious. For example, if the palette and the album probability distributions are different, this joint probability event should be treated separately.

$$H(S_{cp}) = N_c logN_c \times N_p logN_p \tag{3}$$

Similarly, the existence of a correlation between consecutive symbols would indicate a Markov source, so then again should it reduce the entropy and requires reformulations. Besides, if some patterns are somehow restricted to some certain colors (or vice versa), then it should bring in the conditional probabilities, since knowing either of them would increase the expectation and reduce the uncertainty of the other. Suppose that a particular pattern *p* can take some restricted number of colors among the palette, including the color *c*. If there is a conditional probability *p<sub>p</sub>(c)* for the color *c*, then the entropy of this joint event (*p,c*) is 'the uncertainty of *p*' plus 'the uncertainty of *c* when *p* is known', as stated:

$$H(p,c) = H(p) + H_p(c) \wedge H(c,p) = H(c) + H_c(p)$$

However, apparently *H(c)* ≥ *H<sub>p</sub>(c)*, because when *p* and *c* are not independent, then the knowledge of *p* decreases the uncertainty of *c*, or vice versa. Likening that, employing more than 2 colors in a 2-dimensional barcode would make it called *3D*, eke if its symbols are of more than one pattern (i.e. multi-color, multi-pattern alphabet of symbols is in use), then we can name it **4D**. For focusing especially on the colors, it will be assumed in the ensuing discussions that the symbols are of a single pattern (i.e. using different patterns is disregarded), so it is a 3D symbology. (Yet, the formulae can be expended easily towards 4D.) Hence, the color palette equates to the symbol source alphabet, and N is being the palette size. In this case, by holding the previous assumptions, the entropy per symbol can directly be described by *H = N log N* as it was elucidated in (1) previously.

At deduction we can claim that a symbology satisfying all assumptions above provides the largest entropy. So, the formulae on this ground describe the *theoretical maximum* of the information conveying capacity. A symbology design of different properties would cause the entropy decrease, which consequently affects the information density of the code. At the same time, it can also be shown that, over all possible decoding rules, with equiprobable messages, the error probability is minimized [2]; which should be another important motivation of the provided assumptions.

**2.2 Entropy contribution of adding colors**

Increasing the palette size should evidently increase its entropy since *(N+1)log(N+1) > NlogN*. Thus, with the underlying property assumptions, the *maximum entropy contribution* of enlarging palettes (Δ***H*** per symbol) can be generalized as follows:

$$\Delta H(N_2, N_1) = N_2 log N_2 - N_1 log N_1, (N_2 > N_1) \tag{4}$$

When observed the magnitude of this contribution we see that, increasing the palette size from 8 to 14 explicitly uplifts its maximum information conveying capacity by 29.302 bits per symbol. Obviously ΔH is a function of ΔN; yet as well we should also notice that, **Δ*H* is proportional to *N***, since *N<sub>1</sub> log N<sub>2</sub> > (log N<sub>2</sub> – log N<sub>1</sub>)*, and *N<sub>1</sub> log N<sub>2</sub>* increases by *N*, while *(log N<sub>2</sub> – log N<sub>1</sub>)* logarithmically decreases by *N*. This points that higher contributions are attained on larger palette sizes. (For example, ΔH(10,4) = 25.219, while ΔH(14,8) = 29.302, with ΔN = 6.)

In the exclusive comparison, it is natural that the palettes of higher information conveying capacities would be of preference. Once interpreting the information density from an information-theoretical perspective, we can now move forward to the next part inquiring the metrics of the palette evaluation in this means.

## 3 Evaluation metrics for color palettes

Information theory works on contriving optimum coding schemes providing high capacity and authentication performances. In that sense, it is assumed *de facto* that, in the development of robust palettes, the primary goal is to attain the farthest-apart distribution of colors from the reference space for achieving better decoding performances. Also, expanding alphabet sizes (with proper properties) would explicitly increase the entropy, so then the information density of the code. For the sake of higher conveying capacities, the other primary goal is enlarging the palette sizes to the theoretical limits. As it was mentioned previously, comparing palettes of different sizes better attends deciding the optimum number of colors to be included. We can intuitively say that by the way, the process of determining the number of colors should come prior to the color selection.

### 3.1 Comparing palettes of the same size

This comparison is required especially for constructing the best palette of a given number of colors. The methodology serving this requirement is introduced in [9], which suggests a specific metric of *palette quality*. The priority order and the coefficients of every attribute in the palette quality ($Q_p$) calculation are open to adjustment, as the features of the appertaining application could require. The *duality* and the *farthest-duality* concepts were also advised as valuable decision criteria. (For more please cf. [9].)

Evaluating and comparing palettes of different sizes in terms of cost-effectiveness is explored in the following discussions, around the established and formulae.

### 3.2 Cost-effectiveness comparison of palettes of different sizes

Assuming the implementation of the correct construction method assures better designs (if not the best), it can be claimed that the design qualities are to be close to their theoretical limits. When comparing two palettes ($P_1$, $P_2$) of ($N_1$, $N_2$; $N_1 \neq N_2$) is inquired, the ***cost-effectiveness*** metric suggested here for this purpose could provide valuable information. Say the effective density contribution of using $N_2$ versus $N_1$ colors ($N_2 > N_1$) is $\Delta D_c$. The natural cost of boosting information density by adding colors is crowding the source space, consequently reducing the minimum distances. Let the accuracy requirement cost be denoted as $\Delta A_r$. Hence,

$$\Delta D_c(N_2, N_1) = [\, \text{Log}_2(N_2) / \text{Log}_2(N_1) \,] - 1 \tag{5}$$

$$A_r \geq 1 - (\textit{minimum(color differences in palette))} / 765) \tag{6}$$

$$CE(P_2, P_1) = (\Delta D_c - \Delta A_r) / (1 + \Delta A_r) \tag{7}$$

   *CE($P_2$,$P_1$): cost-effectiveness of using $P_2(N_2)$ versus $P_1(N_1)$*

A CE value greater than 0 should indicate an overall gain (e.g. CE < 0 means a loss); so the larger CE is, the larger the gain would be. This metric depicted herein is applied to some developed and HCCB palettes, and the computed results are evaluated. Under the illumination of these discussions, some interesting outcomes are witnessed as the results are studied. We are now ready to share the results and interpretations cultivated from the computations.

## 4 The results

A total of 488 palettes are distinguished for examination, among 2372 palettes constructed in [9], for the number of colors $3 \leq N \leq 100$, all are maximally-separated in the RGB space. The cost-effectiveness comparisons conducted against Microsoft's HCCB palettes are presented in Table 1 (in the order of N). Here we can observe the computed cost-effectiveness values of using some developed palette ($P_2$) instead of HCCB-x ($P_1$), or *vice versa* ($N_2 > N_1$).

Table 1. Cost-effectiveness comparisons against Microsoft's HCCB.

| $P_2$ | $P_1$ | $\Delta D_c$ | $\Delta A_r$ | $CE(P_2,P_1)$ | $P_2$ | $P_1$ | $\Delta D_c$ | $\Delta A_r$ | $CE(P_2,P_1)$ |
|---|---|---|---|---|---|---|---|---|---|
| HCCB4 | 3c | 0,465 | 1,000 | -0,268 | HCCB8 | 3c | 0,893 | 1,000 | -0,054 |
| HCCB4 | 4e | 0,161 | 1,000 | -0,420 | HCCB8 | 4e | 0,500 | 1,000 | -0,250 |
| 6s | HCCB4 | 0,113 | 0,000 | 0,113 | HCCB8 | 5d | 0,292 | 0,332 | -0,030 |
| 7a | HCCB4 | 0,209 | 0,000 | 0,209 | HCCB8 | 6s | 0,161 | 0,000 | 0,161 |
| 8b | HCCB4 | 0,292 | 0,000 | 0,292 | HCCB8 | 7a | 0,069 | 0,000 | 0,069 |
| 9d | HCCB4 | 0,365 | 0,000 | 0,365 | 9d | HCCB8 | 0,057 | 0,000 | 0,057 |
| 10c | HCCB4 | 0,431 | 0,178 | 0,214 | 10c | HCCB8 | 0,107 | 0,178 | -0,060 |
| 11c | HCCB4 | 0,490 | 0,188 | 0,254 | 11c | HCCB8 | 0,153 | 0,188 | -0,030 |
| 12d | HCCB4 | 0,544 | 0,294 | 0,193 | 12d | HCCB8 | 0,195 | 0,294 | -0,077 |
| 13c | HCCB4 | 0,594 | 0,002 | 0,591 | 13c | HCCB8 | 0,233 | 0,002 | 0,231 |
| 14c | HCCB4 | 0,640 | 0,002 | 0,637 | 14c | HCCB8 | 0,269 | 0,002 | 0,267 |
| 15c | HCCB4 | 0,683 | 0,212 | 0,389 | 15c | HCCB8 | 0,302 | 0,212 | 0,075 |

Investigating the Table 1 delivers interesting observations, especially useful to appraise the number of colors to be used. Herein the Microsoft's HCCB-4 and HCCB-8 are coupled with some developed palettes. A 0 value in CE, tells *no gain*, and a negative value tells a *loss*. Notice that, the palette 9d versus HCCB-4 for instance, would provide *4 more* colors (and also *duality*) without any accuracy cost; or 14c instead of HCCB-8 would provide *6 more* colors with almost no accuracy cost. Observations like these should herald that designing efficient palettes of higher information capacities are possible. Ultimately, by studying the results, it could be propounded that, with the correct constructions, it seems feasible to consent some small accuracy costs for the gain of using larger palettes.

## 5 Conclusion

Colors are very efficient sources for information exchange. Including more colors in a palette increases the entropy, yet it could cost declining the reliability of the discrimination. Despite its importance, proper metrics for palette evaluation seem scanty. Here in this study, some explicit definitions and metrics are suggested for capacity evaluation of palettes. The cost-effectiveness metric provides a plausible way of comparing palettes of different number of colors. For the research, the RGB color space is favored, and the color-difference measurement is assumed to be the essential distance criterion. However, corresponding counterparts of these for any different (cubic or not) representation can be derived easily, by valuing analogous approaches. Proposed metric can be applied to axiomatically any distinct number of colors, without any stipulation of the encoding and decoding techniques, applications, devices, operational conditions to work in, or such. Via the comparisons it is proven that, there are designs in which some larger number of colors can be feasibly included in the symbologies with some reasonable or even no accuracy costs, as well as gains. Observations cultivated from the computational results encourage us claiming that, taking aim at better palette designs could be promising; which would help increase the conveying capacity and the inclusive performance of using colors in information representation.